\documentclass[twocolumn,10pt]{asme2e}

\usepackage{graphicx}
\usepackage{xurl}
\usepackage{xcolor}
\usepackage{hyperref}
\usepackage{etoolbox}
\usepackage{orcidlink}
\usepackage{soul}
\usepackage[capitalise]{cleveref}
\usepackage{subcaption}
\confshortname{SSDM2026}
\conffullname{the ASME 2026\\ Aerospace Structures, Structural Dynamics, and Materials Conference}

\confdate{June 8-10}
\confyear{2026}
\confcity{Long Beach}
\confcountry{California}

\papernum{SSDM2026-183328}

\definecolor{preprintbg}{rgb}{0.9,0.9,1}     
\definecolor{preprintlink}{rgb}{0,0,0.5}     
\hypersetup{colorlinks=true,urlcolor=preprintlink,linkcolor=preprintlink,citecolor=preprintlink}
\newcommand{\aamhandle}{https://hdl.handle.net/10016/51138}     
\newcommand{\vorlink}{https://doi.org/10.1115/SSDM2026-183328}  

\newcommand{\preprintbox}{%
  \noindent{\setlength{\fboxsep}{3pt}%
  \colorbox{preprintbg}{\parbox{\dimexpr\textwidth-2\fboxsep\relax}{%
    \raggedright\small
    {\sffamily\bfseries This AAM is available in Green Open Access at:
    {\color{preprintlink}\expandafter\url\expandafter{\aamhandle}}\\
    This is the preprint version of the following article}\\
    Lin Yang, C., Dessena, G., and Bonilla-Manrique, O.~E., ``Affordable Data Collection
    System for UAVs Taxi Vibration Testing,'' \emph{Proceedings of the ASME 2026
    Aerospace Structures, Structural Dynamics, and Materials Conference
    (SSDM2026)}, Long Beach, CA, June 8--10, 2026, Paper SSDM2026-183328.\\
    {\sffamily\bfseries The published version of record can be found at:
    {\color{preprintlink}\mdseries\expandafter\url\expandafter{\vorlink}}}%
  }}}\par\vspace{8pt}%
}

\makeatletter
\patchcmd{\@maketitle}
  {\vspace*{-26pt}}
  {\vspace*{-26pt}\preprintbox}
  {}{\typeout{*** preprintbox patch FAILED ***}}
\makeatother

\title{Affordable Data Collection System for UAVs Taxi Vibration Testing}

\author{Chaoyi Lin Yang\textsuperscript{1}, 
{\tensfb Gabriele Dessena}\textsuperscript{1,}\thanks{Address all correspondence to this author. Email: {gdessena@ing.uc3m.es}.}$\;\;$\orcidlink{0000-0001-7394-9303}, and Oscar E. Bonilla-Manrique\textsuperscript{2} \orcidlink{0000-0003-0541-8310}
    \affiliation{
	\textsuperscript{1}Department of Aerospace Engineering, Universidad Carlos III de Madrid, Leganés, Madrid, Spain\\\textsuperscript{2}Electronic Technology Department, Universidad Carlos III de Madrid, Leganés, Madrid, Spain
    }	
}

\begin{document}

\maketitle    

\begin{abstract}
{\it Structural vibration testing plays a key role in aerospace engineering for evaluating dynamic behaviour, ensuring reliability and verifying structural integrity. These tests rely on accurate and robust data acquisition systems (DAQ) to capture high-quality acceleration data. However, commercial DAQs that provide the required performance and features are often expensive and complex, limiting their accessibility for small-scale research and experimental applications. This work presents the design and experimental validation of an affordable and in-house-developed acceleration DAQ, tested on a small fixed-wing UAV through several Taxi Vibration Test (TVT) runs and ambient vibration measurements. The proposed system integrates several OrangePi 3 LTS single-board computers with multiple LSM6DS3TR-C MEMS inertial measurement units operating simultaneously via an Inter-Integrated Circuit (I2C) communication interface, managed under a Python-based master/slave architecture. Data is acquired at a stable sampling rate of approximately 208 Hz and post-processed using Welch’s method to estimate their Power Spectral Density (PSD). Results confirm the system ability to provide consistent multi-sensor acceleration data and repeatable PSD profiles under the same test conditions; thus, demonstrating its reliability. With a total hardware cost below €600 ($\sim$\$690), the developed DAQ offers a compact, scalable and cost-effective alternative for aerospace vibration analysis and structural testing.}
\end{abstract}

\begin{nomenclature}
\entry{AVT}{Ambient Vibration Tests}
\entry{ANPSD}{Average Normalised PSD}
\entry{CSV}{Comma-Separated Values}
\entry{CI}{Confidence Interval}
\entry{DAQ}{Data Acquisition System}
\entry{FIFO}{First-In, First-Out}
\entry{I2C}{Inter-Integrated Circuit}
\entry{IP}{Internet Protocol}
\entry{IMU}{Inertial Measurement Unit}
\entry{MEMS}{Micro-Electro-Mechanical Systems}
\entry{PSD}{Power Spectral Density}
\entry{SBC}{Single-Board Computer}
\entry{TVT}{Taxi Vibration Test}
\entry{UAV}{Unmanned Aerial Vehicle}
\end{nomenclature}


\section*{INTRODUCTION}

Structural vibration testing plays a key role in aerospace engineering, as it enables the evaluation of dynamic behaviour, the identification of structural weaknesses, and the verification of compliance with safety and performance requirements. Aerospace structures are routinely exposed to complex dynamic loads during operation, making vibration analysis an essential tool for ensuring reliability throughout the design and testing phases \cite{Dessena2025a}. Among the various experimental techniques applied in this context, vibration testing provides critical insight into structural response under realistic excitation conditions.

To conduct vibration tests effectively, accurate and reliable data acquisition systems (DAQs) are required. These systems serve as the interface between physical phenomena and digital analysis, capturing acceleration signals from sensors and converting them into data suitable for post-processing and frequency-domain evaluation \cite{Rainieri2014}. In aerospace applications, DAQ systems are commonly used in procedures such as Ground Vibration Tests, Taxi Vibration Tests (TVTs), and Ambient Vibration Tests (AVT), where high-quality acceleration data are required for identifying dominant vibration modes and validating structural models.
However, commercial DAQ solutions capable of meeting aerospace testing requirements are often characterised by high cost, complex architectures, and limited accessibility for small-scale research, experimental platforms, or educational environments. This economic and practical barrier has motivated growing interest in the development of alternative low-cost data acquisition solutions that can still deliver reliable and repeatable measurements \cite{Payet2025}. Advances in Single-Board Computers (SBC) and Micro-Electro-Mechanical Systems-(MEMS-)based inertial sensors have opened new opportunities for developing compact and scalable DAQ systems tailored to experimental vibration analysis.

In this context, this work presents the design and experimental application of an in-house-developed, affordable acceleration data acquisition system intended for aerospace vibration testing applications. The proposed system integrates a SBC with multiple MEMS inertial measurement units operating concurrently under a Python-based master–slave software architecture. The system is used in TVTs and AVTs conducted on a small fixed-wing Unmanned Aerial Vehicle (UAV), with results analysed in the frequency domain using Power Spectral Density (PSD) estimation. With a total hardware cost below €600 ($\sim$\$690), the developed DAQ system demonstrates a practical and cost-effective alternative for structural vibration testing while maintaining performance and repeatability comparable to more expensive commercial solutions.
\section*{SYSTEM ARCHITECTURE}
The proposed data acquisition system is designed to provide a flexible and scalable solution for vibration testing in aerospace applications, while maintaining a low overall cost. The system is conceived to provide a compact, scalable, and cost-effective alternative to commercial DAQ solutions while maintaining sufficient performance for structural dynamics analysis.

The overall architecture follows a distributed master–slave configuration, in which multiple acquisition nodes are responsible for local sensor management and data collection, while a central master unit coordinates the system and performs data aggregation and storage. This approach allows the system to be easily expanded or reduced depending on the number of required measurement points, without altering the core functionality.

\subsection*{Hardware Architecture}
The core of the DAQ system consists of an Orange Pi 3 LTS SBC\footnote{\url{https://www.orangepi.org/html/hardWare/computerAndMicrocontrollers/details/orange-pi-3-LTS.html}} acting as the processing unit for each acquisition module, interfaced with multiple Adafruit LSM6DS3TR-C 6 degrees-of-freedom (DoFs) inertial measurement units (IMUs)\footnote{\url{https://www.adafruit.com/product/4503?srsltid=AfmBOopFqvZNSULVXQAspbTo7Vfuk5xISjZ-4G_q_H2cUSwbhKb1mKul}}. These sensors are selected due to their compact size, low power consumption, and configurable output data rates and measurement ranges, which make them cost-effective devices for capturing vibration data.

Each acquisition module is responsible for collecting acceleration data locally and transmitting it to a central master unit. This modular approach allows additional acquisition nodes to be integrated into the system without altering the overall architecture, supporting scalability in the number of measurement points.

\subsection*{Master-Slave Layout}
The DAQ system follows a distributed master–slave layout, in which multiple slave modules operate under the control of a central master unit. Each slave module is responsible for configuring its connected sensors and executing the data acquisition process. This distributed approach reduces the computational burden on the master unit and allows data collection to occur in parallel across multiple structural locations.

The master unit serves as the coordination and control node of the system. It broadcasts configuration parameters—such as sampling frequency, test duration, and sensor range—to all slave modules, ensuring consistent acquisition settings across the entire sensor network. In addition, the master ensures synchronised acquisition conditions across multiple measurement points, so that the data acquired are consistent throughout the structure. During the acquisition process, each slave module continuously transmits the acquired data to the master unit, enabling real-time aggregation and monitoring while acquisition is ongoing. It is worth to mention that the number of sensors which connects to the slave modules can be tailored by the operator, enabling flexibility on the system.

\subsection*{Sensor Network and Communication Strategy}
Acceleration measurements are obtained using MEMS-based IMU sensors operating as accelerometers. 
Multiple sensors are connected to each slave module through an Inter-Integrated Circuit (I2C) communication interface integrated in the SBCs. A key challenge in multi-sensor configurations using I2C communication lies in address conflicts, as multiple identical sensors typically share the same fixed I2C address. To overcome this limitation, an I2C multiplexer\footnote{\url{https://www.adafruit.com/product/5626?srsltid=AfmBOopF-JjX1PYElc8RmOrpO7GuzKl8j91ktRMehvdPrDyEVPCRXMGD}} was introduced between the SBC and the IMU sensors. The multiplexer enables the SBC to selectively prioritise individual sensor channels, effectively isolating sensors on separate I2C buses while maintaining a single physical interface.

\begin{figure*}[!ht]
\centering
\includegraphics[width=0.7\textwidth]{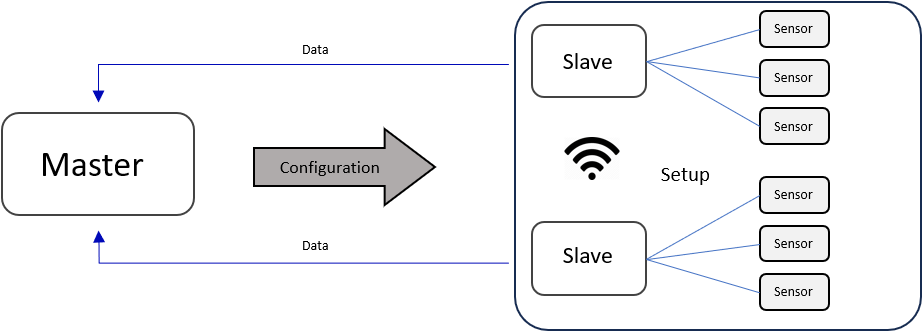}
\caption{SYSTEM ARCHITECTURE BLOCK DIAGRAM}
\label{fig:fig1} 
\end{figure*}

This solution is chosen to allow the use of multiple identical IMUs without requiring hardware address modification or additional communication interfaces. The sensor network design prioritises scalability and ease of installation. Additional sensors can be incorporated by extending the multiplexer channels or adding further slave modules, allowing the system to adapt to different test configurations and structural layouts. The final configuration considered in this work includes a master and two slave modules, communicating with four IMUs each transmitting acceleration data in the $x$, $y$, and $z$ axes. Communication between the master and slave nodes is carried out using Internet Protocol (IP) over wireless links, i.e. WiFi. The proposed system architecture is shown in \cref{fig:fig1}.

\section*{SOFTWARE CONFIGURATION}
Data acquisition is implemented through a distributed software framework designed to ensure stable sampling, synchronised configuration, and reliable multi-sensor operation. Each slave module manages a set of MEMS accelerometers connected via an I2C bus and is responsible for sensor initialisation, data reading, and local buffering. Acquisition parameters, including sampling frequency, measurement range, and acquisition duration, are defined by the master unit and broadcast to all slave modules prior to the start of each test. Once configured, slave modules perform continuous sensor polling: Acceleration data are read directly from the sensor registers and timestamped locally to preserve temporal consistency across channels.

During operation, data acquisition and transmission occur concurrently. Rather than waiting for the completion of the test, slave modules stream acquired data to the master unit while acquisition is ongoing. This approach reduces memory usage at the slave level, enables real-time monitoring of data integrity, and minimises the risk of data loss during long-duration tests. Local buffering is maintained only as required to ensure reliable data transfer under variable communication conditions. The master unit manages global test control, receives incoming data streams from all slave modules, and organises the datasets for subsequent post-processing. The master–slave architecture allows acquisition tasks to be executed in parallel across multiple structural locations, reducing computational bottlenecks and enabling scalable deployment without modifying the core software structure.

An alternative acquisition strategy based on sensor First-In, First-Out (FIFO) buffering \cite{Fedasyuk2019} was evaluated during development, but ultimately not implemented. While FIFO-based acquisition can reduce bus transactions, it introduces additional complexity in data alignment and timing control within the Python-based framework. The adopted direct polling approach was therefore selected as a more reliable solution for ensuring consistent timing behaviour across multiple sensors. After reaching the master module, data is saved in a comma-separated values (CSV) file to be readily exported for any post-processing necessary. 

Although the use of Python introduces execution overhead, experimental results demonstrate that the implemented methodology provides sufficient timing stability and frequency resolution for typical aerospace vibration testing scenarios.

\section*{EXPERIMENTAL SETUP AND PROCEDURE}
Experimental validation of the proposed DAQ system is conducted using a small fixed-wing UAV as the test structure: the Volantex RC Ranger 2400\footnote{\url{https://www.volantexrc.eu/volantex-rc-ranger-2400-professional-fpv-uav-carrier-757-9-kit-p-307.html}} (from now on referred to as the \textit{Ranger}). The Ranger is a 2.4 m wingspan drone with an overall length of 1.23 m and a total mass (battery excluded) of 1.7 kg. The objective of the experimental test is to evaluate the system capability to acquire consistent and repeatable acceleration data under representative (AVT and TVT) vibration conditions, while operating in a multi-sensor configuration.

\begin{figure*}[!ht]
\centering
\includegraphics[width=.7\textwidth]{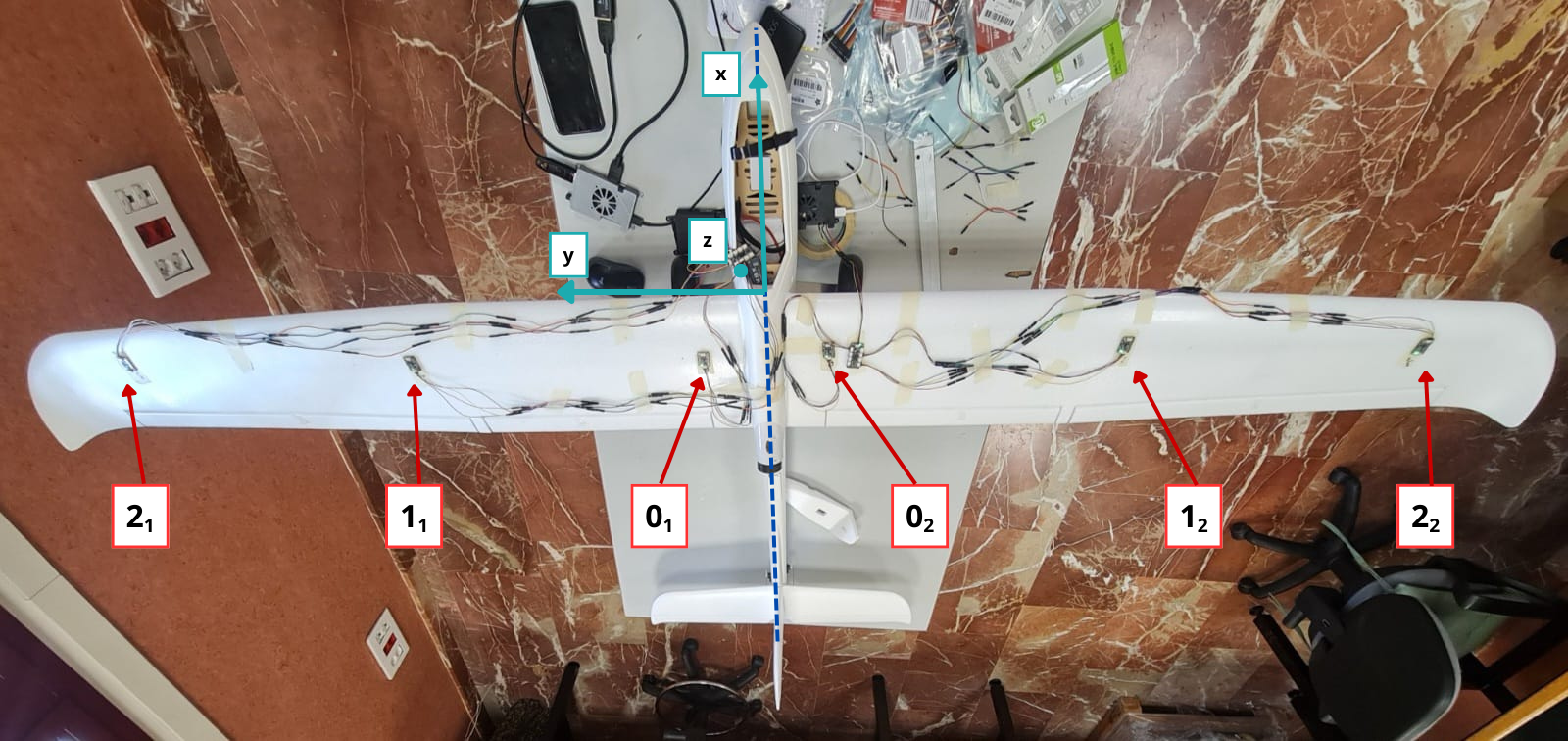}
\caption{VOLANTEX RC RANGER 2400 INSTRUMENTED WITH THE PROPOSED LOW-COST DAQ.}
\label{fig:fig2} 
\end{figure*}

The DAQ system is installed directly on the Ranger structure, with multiple MEMS accelerometers distributed evenly along the wing outer skin to capture spatial variations in structural response, as shown in \cref{fig:fig2}. The sensors are placed on the vertical projection of the wing spar at predefined locations, using a symmetric layout to ensure comparable measurements across the structure. Each wing is instrumented with an independent slave acquisition module, allowing parallel data collection from multiple measurement points. On each slave module, each I2C bus on the multiplexer is defined by its respective number and the sensors are wired according to the name assignation. The position of the six accelerometers is shown in \cref{fig:fig2} and reported in \cref{tab:tab1}. Note that only six accelerometers are used due to the unavailability of enough connection cables to accommodate eight sensors on the wing at the time.

\begin{table}[!ht]
\caption{SENSOR POSITIONS.\label{tab:tab1}}
\begin{center}
\label{table_sensors}
\begin{tabular}{c c c}
\hline
Sensor No. & $x~[\mathrm{cm}]$ & $y~[\mathrm{cm}]$ \\
\hline
$0_1$ & $-8.6$ & $8.6$ \\
$1_1$ & $-8.6$ & $52.6$ \\
$2_1$ & $-8.6$ & $97.6$ \\
$0_2$ & $-8.6$ & $-8.6$ \\
$1_2$ & $-8.6$ & $-39.2$ \\
$2_2$ & $-8.6$ & $-96.6$ \\
\hline
\end{tabular}
\end{center}
\end{table}

\begin{figure}[!ht]
\centering
	\begin{subfigure}[t]{.45\textwidth}
	\centering
		{\includegraphics[width=\textwidth,keepaspectratio]{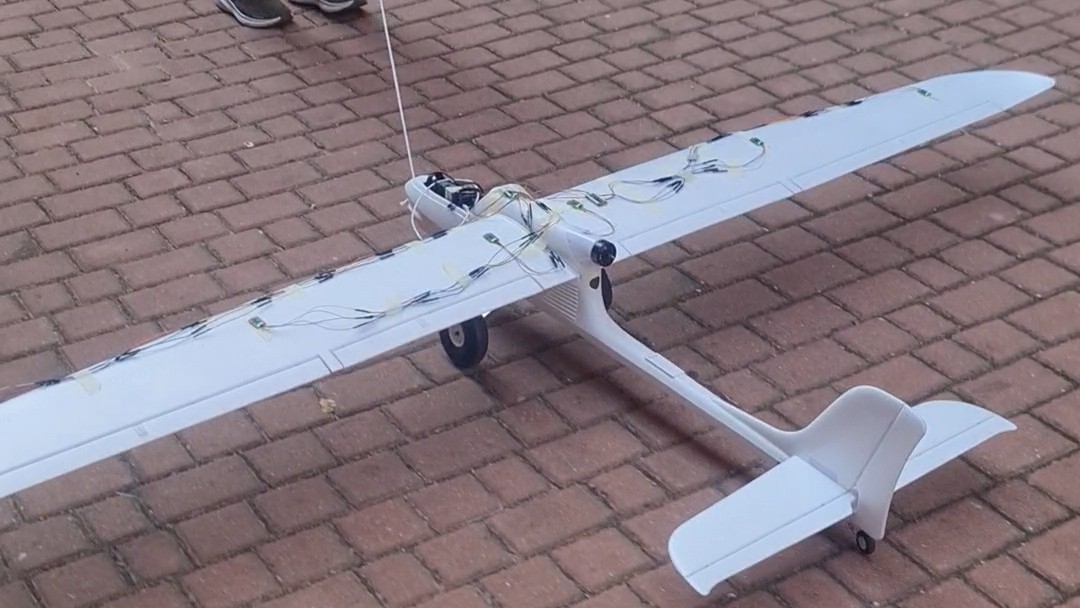}}
		\captionsetup{font={it},justification=centering}
		\subcaption{\label{fig:fig3a}}	
	\end{subfigure}
    \begin{subfigure}[t]{.45\textwidth}
	\centering
		\includegraphics[width=\textwidth,keepaspectratio]{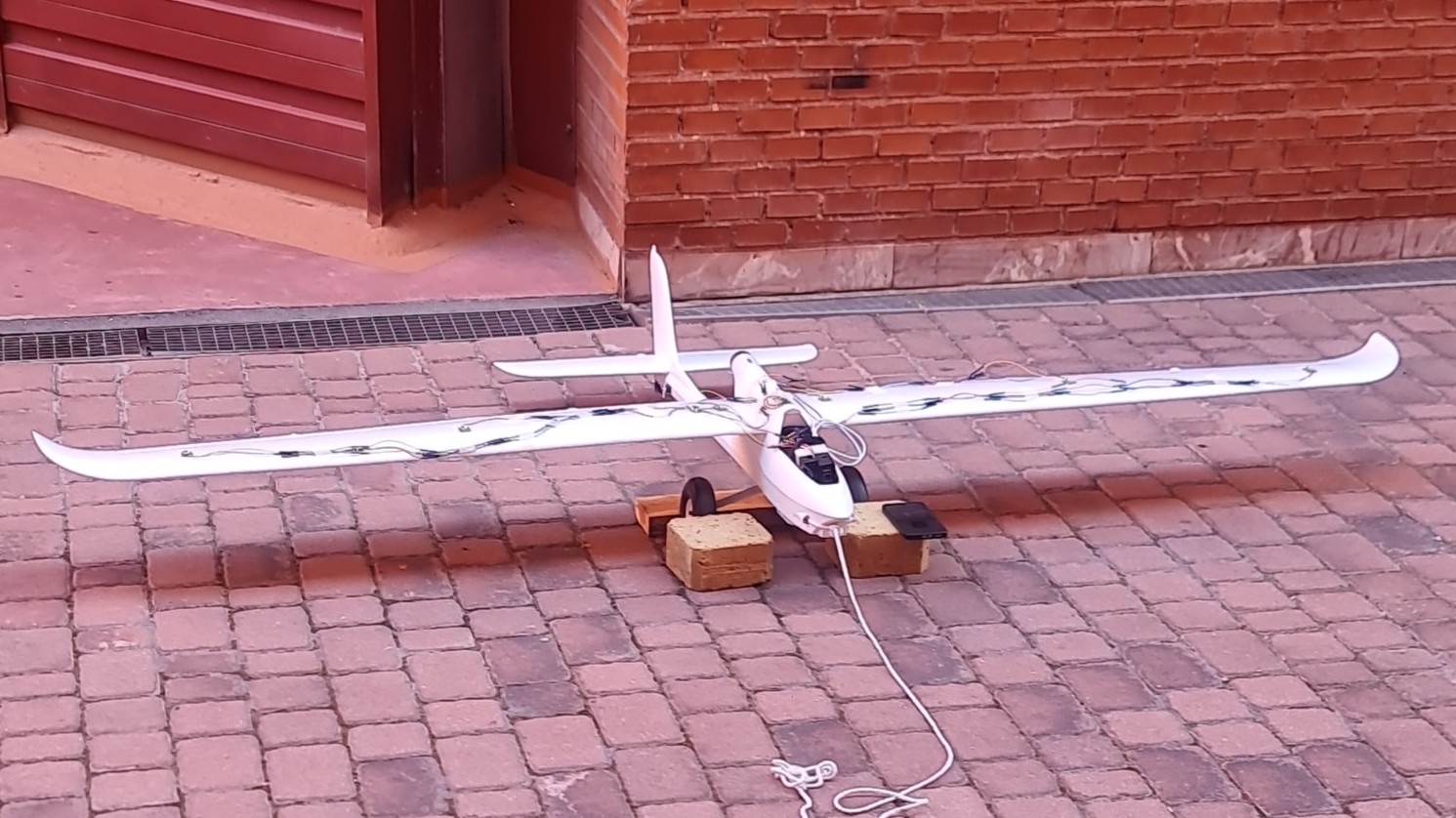}
		\captionsetup{font={it},justification=centering}
		\subcaption{\label{fig:fig3b}}	
	\end{subfigure}
	\caption{RANGER DURING (a) TVT AND (b) AVT.}
	\label{fig:fig3}
\end{figure}

Two types of vibration tests are performed: TVTs and AVTs. During the former the Ranger is subjected to ground-induced excitations while being towed along an uneven surface (\cref{fig:fig3a}), while in the latter, its wheels are secured with wooden blocks to constrain any movement (\cref{fig:fig3b}). The duration of each TVT test is approximately 60 s, which is configured prior to the start of each test. The TVT test provides broadband excitation representative of operational ground conditions. The second consists of AVTs, conducted with the Ranger at rest, where only low-level environmental excitations are present. The duration of the AVT test is 20 minutes, longer than that of the TVT, to ensure the capture of the vibration data. All data is sampled at $f_s =$ 208 Hz from six tri-axial accelerometers, accounting for a total of 18 independent measurement channels. 

This combination of tests enables the assessment of system performance in both forced and ambient excitation scenarios. During each test, the acquisition parameters are configured centrally and applied uniformly across all slave modules to maintain consistent measurement conditions. Acceleration data are acquired continuously throughout the duration of the test and transmitted to the master unit in parallel with ongoing acquisition.

\subsection*{Data Processing}

The acceleration data acquired during the vibration tests are post-processed to characterise the dynamic response of the structure in the frequency domain. All recorded datasets are stored in CSV format and processed offline to ensure consistency and repeatability across tests. The primary analysis tool employed is the PSD, which provides a frequency-domain representation of the vibration energy distribution. PSD estimation allows dominant frequency components to be identified and compared across different test conditions, making it particularly suitable for structural vibration analysis and validation purposes.

In particular, Average Normalised PSD (ANPSD) estimates are considered in this work, which allow for a better visualisation of amplitude peaks in the signal spectra \cite{Felber1993}. This approach reduces variance in the spectral estimate while maintaining adequate frequency resolution, making it well-suited for experimental vibration data with stochastic characteristics. Prior to spectral computation, the acceleration signals are adjusted to compensate any constant bias introduced by sensor offsets. A Hamming window is applied to each segment to minimise spectral leakage. The segment length and overlap ratio are selected to provide a balance between frequency resolution and noise reduction, ensuring consistent processing across all datasets.
In addition, confidence intervals (CIs) of 95\% are computed to assess the statistical consistency of the estimated spectra. 

\section*{RESULTS}

The proposed low-cost DAQ system is evaluated through a series of experimental vibration tests, previously described, conducted on a small fixed-wing UAV. Two complementary test scenarios are considered: TVTs, representing externally excited conditions, and AVTs, representing low-energy, operational-at-rest conditions. The acquired acceleration data are analysed in the frequency domain to assess consistency, repeatability, and suitability for structural vibration characterisation.

\subsection*{TVT Results}
\begin{figure}[!ht]
\centering
\includegraphics[width=\columnwidth]{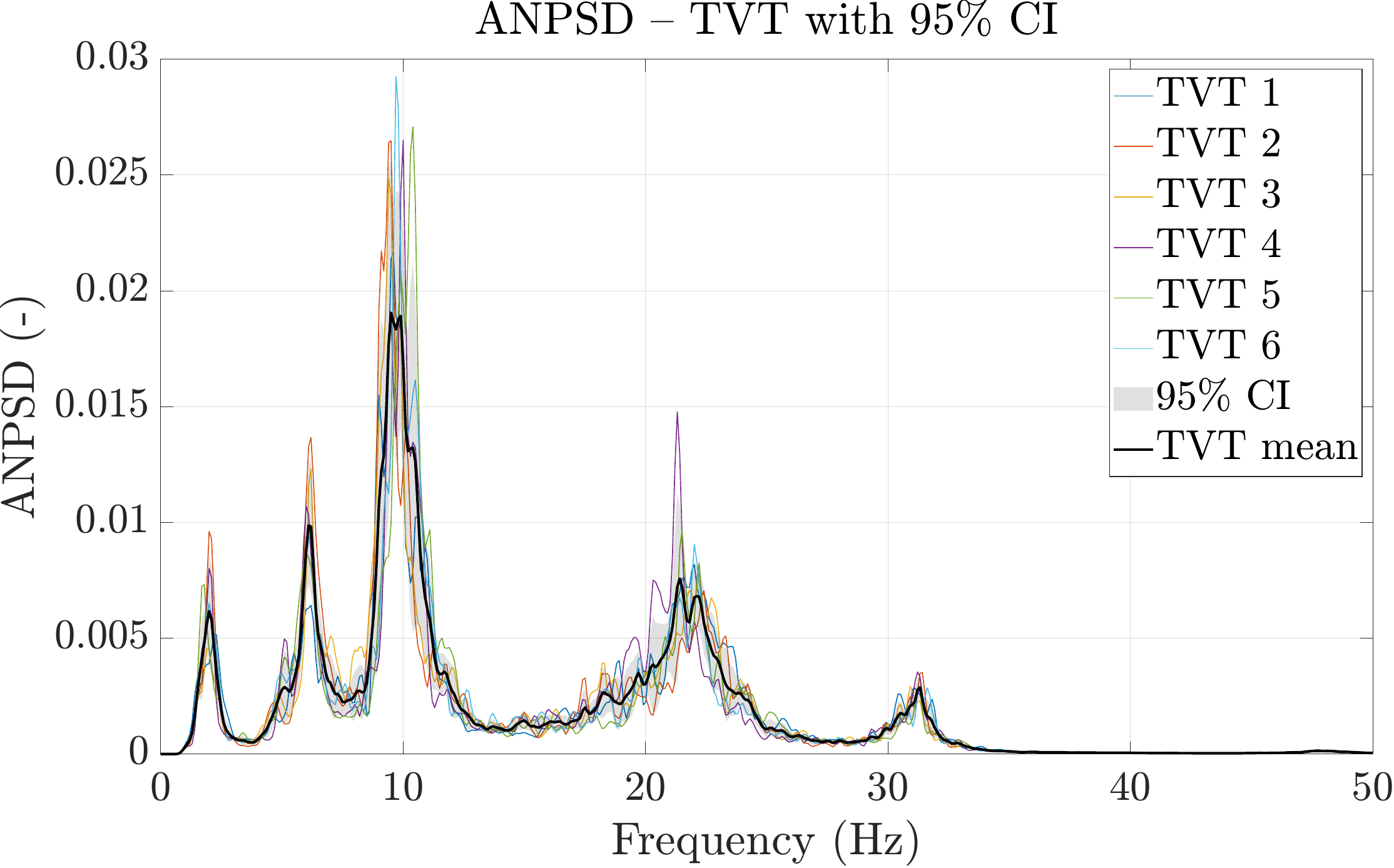}
\caption{ANPSDS OF TVTS.}
\label{fig:fig4} 
\end{figure}

The TVT generates broadband excitation due to ground interaction and external disturbances. The ANPSD estimates, shown in \cref{fig:fig4}, are obtained from multiple, six, TVT runs and exhibit consistent spectral shapes, with dominant vibration energy concentrated at low frequencies. 
Close-matching PSD profiles indicate stable acquisition behaviour and repeatability across measurement points. Nevertheless, minor variations in amplitude between individual runs can be attributed to unavoidable differences in excitation levels and ground surface irregularities. Nevertheless, the overall frequency content remains consistent, demonstrating that the DAQ system is capable of capturing repeatable vibration signatures under dynamic excitation conditions. These results confirm the suitability of the proposed system for vibration testing scenarios involving non-stationary and externally driven inputs.

\subsection*{AVT Results}
\begin{figure}[!ht]
\centering
\includegraphics[width=\columnwidth]{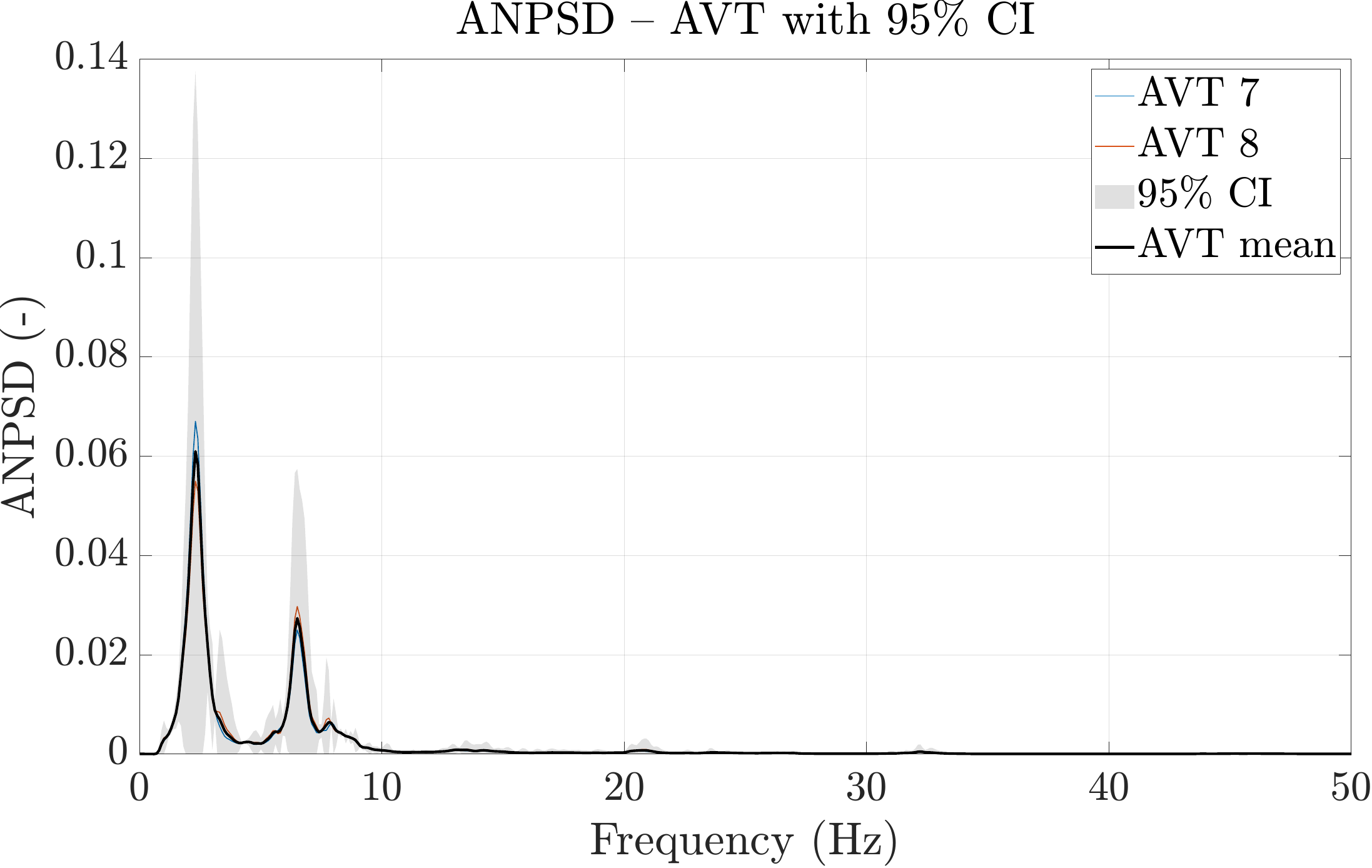}
\caption{ANPSDS OF AVTS.}
\label{fig:fig5} 
\end{figure}

AVTs are conducted with the Ranger at rest, capturing the structural response under low-level environmental excitation. Although not shown in the ANPSD of \cref{fig:fig5}, the resulting PSD profiles show significantly lower overall energy compared to TVT results, with dominant frequencies appearing at very slightly lower amplitudes. Despite the reduced excitation, the system successfully captures coherent frequency content across all sensors.
Two AVT runs are conducted, returning consistent measurements between repetition. This indicates that the DAQ system maintains stable performance even under low signal-to-noise conditions. This behaviour is particularly relevant for structural dynamics applications, where ambient vibration measurements are often used to identify natural frequencies without external forcing.

\section*{Comparison Between TVT and AVT}

A direct comparison between TVT and AVT results highlights the sensitivity of the acquisition system to different excitation regimes. TVT measurements exhibit higher energy levels and broader frequency content due to continuous external excitation, whereas AVT measurements reflect the intrinsic dynamic characteristics of the structure under minimal excitation.

\begin{figure}[!ht]
\centering
\includegraphics[width=\columnwidth]{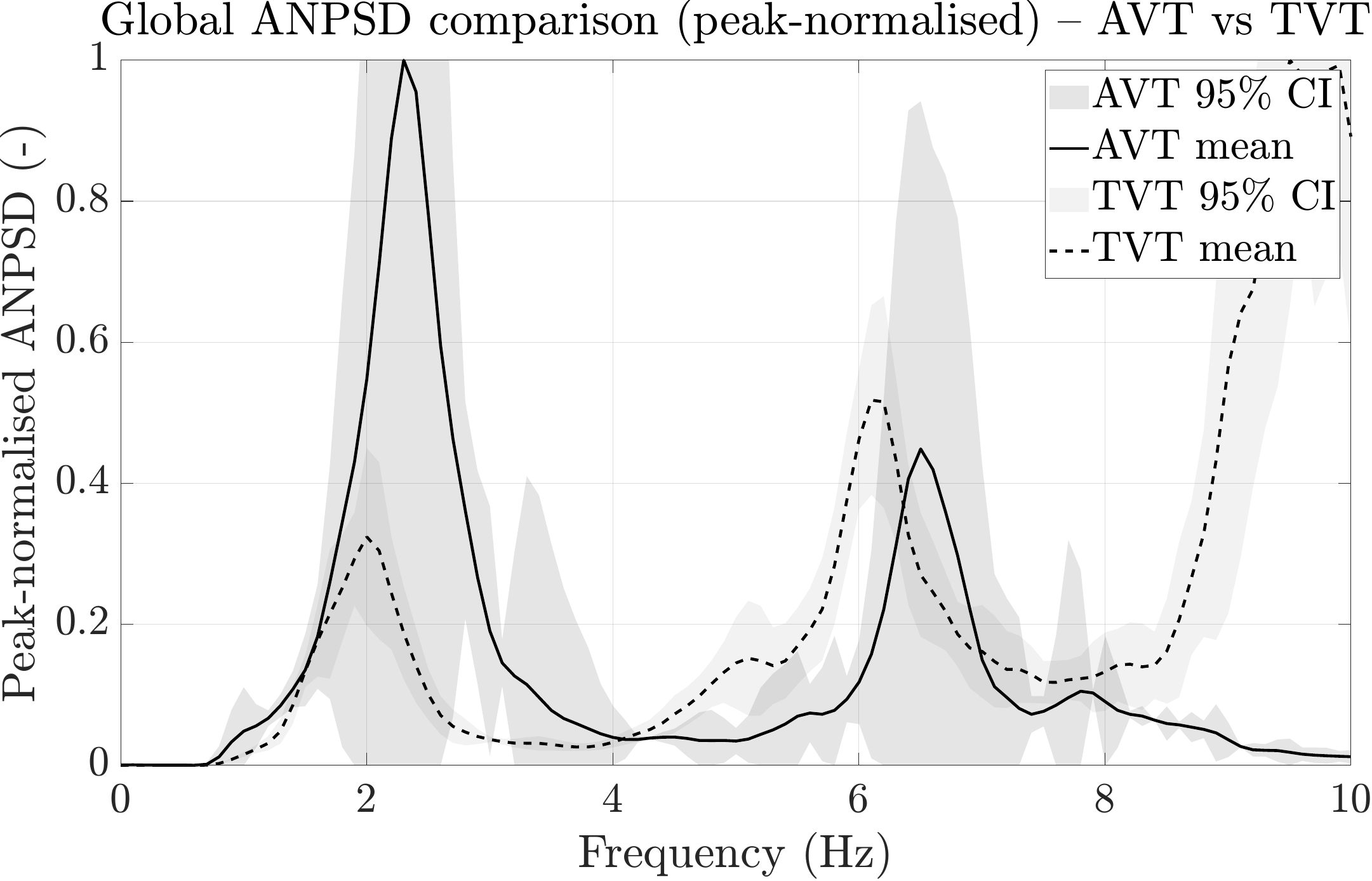}
\caption{PEAK-NORMALISED ANPSD WITH CI OF 95\%}
\label{fig:fig6} 
\end{figure}

Notably, five dominant peaks are observed in the TVT ANPSDs in \cref{fig:fig4}. These peaks are also visible in \cref{fig:fig5}; however, the amplitudes of the last three modes are extremely small in the AVT case. To enable a fair comparison, the peak-normalised mean ANPSDs of AVTs and TVTs, over the frequency interval from 0 to 10 Hz, are therefore shown in \cref{fig:fig6}. This interval is selected as it contains the two highest-amplitude modes of the less excited case, namely AVT.
The first two peaks frequency is consistent for both TVT and AVT. Although, a small shift is observed between the two cases. This is physically consistent with the test conditions, different input amplitude for a rather flexible system, and confirm that the measured responses are governed by the excitation environment rather than artifacts of the acquisition system. This comparison further validates the ability of the DAQ system to reliably distinguish between different vibration scenarios.

\subsection*{Repeatability of the system}
To assess measurement repeatability, a statistical analysis of multiple test runs is performed using ANPSDs and 95\% CI, as shown in \cref{fig:fig4,fig:fig5}. The narrow confidence bounds observed at lower frequencies indicate a high level of consistency across repeated measurements, whereas wider intervals at higher frequencies reflect increased variability. Clearly, the CIs associated with the AVTs are larger than those of the TVTs. This behaviour is attributed to the lower number of test repetitions, namely two for AVTs compared to six for TVTs, as well as to the lower excitation amplitude characterising the AVT configuration.

Overall, the results demonstrate that the proposed DAQ system provides reliable and repeatable acceleration measurements across multiple sensors and test conditions. The consistency of the spectral content observed across repeated experiments confirms that the system performance is comparable, in practical terms, to that required for experimental vibration analysis in aerospace applications.

\section*{CONCLUSION}
Low-cost and flexible instrumentation solutions are increasingly relevant in experimental aerospace engineering, particularly for small-scale research, prototyping, and educational applications. In this context, the developed data acquisition system demonstrates that reliable vibration measurements can be achieved without relying on high-cost commercial Data Acquisition System (DAQ) platforms.

The total hardware cost of the proposed system remains below €600 ($\sim$\$690), including 3x single-board computers, 8x micro-electro-mechanical systems inertial sensors, 2x multiplexers, and auxiliary components. This cost level is significantly lower than that of conventional aerospace-grade DAQs, which often require substantial financial investment and complex installation procedures. The compact size and modular architecture further simplify deployment on lightweight platforms such as Unmanned Aerial Vehicles (UAV), where space, weight, and accessibility are critical constraints. Beyond cost reduction, the system exhibits a high degree of scalability and adaptability. The distributed master–slave architecture allows additional acquisition modules to be incorporated or removed according to the specific requirements of each test campaign, without changes to the core software structure. This flexibility makes the system suitable for a wide range of experimental scenarios, including preliminary vibration surveys, structural dynamics studies, and repeated testing during iterative design processes.

Experimental results obtained from taxi and ambient vibration tests confirm that the system provides consistent and repeatable acceleration measurements across multiple sensors. Despite inherent limitations associated with the use of Python and I2C-based (Inter-Integrated Circuit) communication—such as reduced maximum sampling rates compared to specialized hardware—the achieved performance is sufficient to characterise the dominant frequency content of typical aerospace vibration phenomena. The observed repeatability and stability of the spectral responses indicate that the proposed DAQ system can deliver data quality comparable, for many applications, to that of more expensive commercial solutions.

Overall, the presented system represents a practical and cost-effective alternative for aerospace vibration testing, closing the gap between low-cost prototyping platforms and professional experimental instrumentation. Future developments may focus on improving timing precision, increasing achievable sampling rates through optimised software and/or routines, extending the system to multi-axis and gyroscopic measurements, and complete its validation against industry standard equipment.

\bibliographystyle{asmems4}
\begin{acknowledgment}
The authors would like to thank Dr Francesco M.A. Mitrotta from the Department of Aerospace Engineering at the Universidad Carlos III de Madrid for helping out running the hereby reported testing campaign and providing the Volantex RC Ranger 2400 used.
This work has been supported by the Madrid Government (\textit{Comunidad de Madrid}, Spain) under the Multiannual Agreement with the UC3M (IA\_aCTRl-CM-UC3M).
\end{acknowledgment}
%
% ...


\end{document}